\documentclass[fleqn,usenatbib]{mnras}

\usepackage{newtxtext,newtxmath}
\usepackage[T1]{fontenc}
\usepackage{ae,aecompl}

\usepackage{graphicx}	% Including figure files
\usepackage{amsmath}	% Advanced maths commands
\usepackage{cprotect}

\usepackage{caption}
\usepackage{subcaption}
\usepackage{hyperref}

\newcommand{\pc}{\,{\rm pc}\,}% parsec
\newcommand{\kpc}{\,{\rm kpc}\,}% kiloparsec
\newcommand{\Myr}{\,{\rm Myr}\,}% megayears
\newcommand{\Gyr}{\,{\rm Gyr}\,}% gigayears
\newcommand{\kms}{\,{\rm km s}$^{-1}$\,}% kilometres per second
\newcommand{\G}{\,{\rm G}\,}% gauss
\newcommand{\muG}{\,$\mu$\G\,}% microgauss
\newcommand{\erg}{\,{\rm erg}\,}% gram

\newcommand{\emf}{\mbox{\boldmath ${\cal E}$} {}}
\newcommand{\mean}[1]{\overline{#1}}

\newcommand{\eref}{Eq.~\,\ref}% referring equations
\newcommand{\fref}{Fig.~\,\ref}% referring figures
\newcommand{\sref}{Sec.~\,\ref}% referring sections
\newcommand{\tref}{Table~\,\ref}% referring Tables
\let\ACMmaketitle=\maketitle
\renewcommand{\maketitle}{\begingroup\let\footnote=\thanks \ACMmaketitle\endgroup}

\title[Cosmic Ray Dynamo]{On the Combined Role of Cosmic Rays and Supernova-Driven Turbulence for Galactic Dynamos\footnote{We dedicate this manuscript to the memory of Prof. Karl-Heinz R\"adler (1935-2020).}}

\author[A. Bendre]{
Abhijit B. Bendre,$^{1}$\thanks{E-mail: abhijit@iucaa.in}
Detlef Elstner$^{2}$\thanks{E-mail: elstner@aip.de}
and
Oliver Gressel$^{2}$\thanks{E-mail: ogressel@aip.de}
\\
$^{1}$IUCAA, Post Bag 4, Ganeshkhind, Pune 411007, India\\
$^{2}$Leibniz-Institut f\"ur Astrophysik Potsdam (AIP), An der Sternwarte 16, 14482, Potsdam, Germany
}

\date{Accepted XXX. Received YYY; in original form ZZZ}

\pubyear{2015}

\begin{document}
\label{firstpage}
\pagerange{\pageref{firstpage}--\pageref{lastpage}}
\maketitle

\begin{abstract}
Large-scale coherent magnetic fields observed in the 
nearby galaxies are thought to originate by a mean-field 
dynamo. This is governed via the turbulent electromotive 
force (EMF, $\overline{\emf} $) generated by the helical 
turbulence driven by supernova (SN) explosions in the 
differentially rotating interstellar medium (ISM). In 
this paper we aim to investigate the possibility of 
dynamo action by the virtue of buoyancy due to a cosmic
ray (CR) component injected through the SN explosions. 
We do this by analysing the magnetohydrodynamic
simulations of local shearing box of ISM, in which the
turbulence is driven via random SN explosions and the
energy of the explosion is distributed in the CR and/or
thermal energy components. We use the magnetic field
aligned diffusion prescription for the propagation of 
CR. We compare the evolution of magnetic fields in the 
models with the CR component to our previous models 
that did not involve the CR. We demonstrate that the 
inclusion of CR component enhances the growth of dynamo
slightly. We further compute the underlying dynamo 
coefficients using the test-fields method, 
and argue that the entire evolution of the large scale 
mean magnetic field can be reproduced with an 
$\alpha-\Omega$ dynamo model. We also show that the 
inclusion of CR component leads to an unbalanced 
turbulent pumping between magnetic field components
and additional dynamo action by the R\"adler effect.
\end{abstract}


%%%%%%%%%%%%%%%%%%%%%%%%%%%%%%%%%%%%%%%%%%%%%%%%%%

%%%%%%%%%%%%%%%%% BODY OF PAPER %%%%%%%%%%%%%%%%%%

\section{Introduction}
\label{sec:intro}
The origin of kilo-parsec scale magnetic fields observed in 
the nearby galaxies through the polarized radio synchrotron 
emission \citep[eg.][etc.]{fletcher_nearby_2010,Beck2012} is 
attributed to the large scale dynamo operating in the ISM. 
This is driven mainly via the helical turbulent motions in 
the interstellar medium, coupled with the differential shear 
and vertical density stratification. This mechanism, along 
with some phenomenological approximations about the properties 
of background turbulence, in principle explains the growth of 
magnetic fields from small initial strengths to large-scale 
equipartition strengths against the diffusive losses 
\citep{beck_1996,shukurov_2005,beck_wielebinski}, and the 
characteristic times it takes for the field to reach the 
equipatition strength turn out to be of the order of $\sim 
\Gyr$. This is perhaps a much too slow to account for the 
strong equipartition strength magnetic fields observed 
in the high redshift galaxies with $z>1$ \citep[eg.][]{2008Natur.454..302B} 
or even for that in the slowly rotating nearby galaxies. This 
discrepancy leads one to invoke some additional mechanism 
such as cosmic rays boosting the typical dynamo action. 
The idea of CR driven dynamo was initially discussed by 
\cite{parker1992}, this predicted the possibility of 
enhanced dynamo action by the virtue of additional CR 
buoyant instability, that inflates the magnetic field 
structures \citep[see also][]{axel2018}. Based on the 
conventional dynamo formulation Parker further suggested 
a simple model for the flux loss through the gaseous disc 
due to buoyancy by substituting the transport terms 
$B_\phi/t_d$. These terms are supposed to encapsulate 
the non advective flux transport associated with the 
buoyant instability, and leads to the fast dynamo action 
in characteristic field mixing times. \cite{Hanasz2000} 
indirectly verified such a dynamo action via the numerical
simulations of rising magnetic 
flux tubes and found e-folding times of mean field of 
the order of $100\Myr$. Supplementing this 
\cite{Hanasz2009,seijkowski_09,kulpa2015,Girichidis2016} etc. also
demonstrated the fast amplification of regular magnetic 
fields via the direct MHD simulation of global galactic 
ISM including cosmic ray driven turbulence, along with 
the differential shear (but excluding the viscous term). 
To complement this, we aim here to extend our previous 
analysis of dynamo mechanism in SN driven ISM turbulence 
\cite{bendre_15}, by including the CR component and 
investigate the influence of magnetic field dependent 
propagation of CR on the dynamo. Here we focus on 
estimating the dynamo coefficients from the direct 
MHD simulations and effect CR component has on them by 
comparing with our previous analysis without the CR.

This manuscript is structured as follows: In \sref{sec:model_equations} we describe our simulation 
setup with relevant equations, followed by a description 
of simulated models and parameters in
\sref{sec:model_parameters}. In \sref{sec:results} we discuss 
the overall outcomes of the simulations, estimation of dynamo
coefficients and compare the results with that in our previous
analysis in \cite{bendre_15}. In \sref{sec:conclusions} we list
out main conclusion, which is followed a summary
in \sref{sec:summary}.

\section{Model Equations}
\label{sec:model_equations}
We use the NIRVANA MHD code \citep{nirvana} to simulate 
our system of turbulent ISM in a shearing Cartesian local 
box of dimensions $0.8 \kpc \times 0.8  \kpc $ in $x$ 
(radial) and $y$ (azimuthal) direction and $-2.13 \,{\rm 
to}\,2.13 \kpc$ in $z$ (vertical) direction, with galactic 
midplane situated at $z =0$. We use the shearing periodic 
boundary conditions in $x$ direction to incorporate the 
effect of differential shear, along with angular velocity 
$\Omega$ that scales as $R^{-1}$ (where $R$ is the radius
) with $\Omega = 100$\kms$\kpc^{-1}$ at the center of the box, mimicking the flat rotation 
curve. In the azimuthal $y$ direction we use periodic 
boundary conditions. While at the $z$ boundaries we allow 
the gas outflow, by setting the inward velocity components 
to zero, while for the CR energy we use the gradient 
condition at the $z$ boundaries. With this setup we solve 
the following set of differential equations, 
\begin{eqnarray}
      \frac{\partial\rho}{\partial t} + \nabla\cdot\left(\rho \mathbf{U}\right)
	     & = & 0\,, 
	     \nonumber\\ \nonumber\\
      \frac{\partial\left(\rho\mathbf{U}\right)}{\partial t} 
           + \nabla\cdot\left[\rho\mathbf{UU}+\mathrm{p}^{\star}
           - \mathbf{BB}\right] & = &
	   - 2 \rho\ \Omega\ \hat{z}\times\mathbf{U} \nonumber \\ & &
           + 2\rho\ \Omega^{2} q x \hat{x} \nonumber\\ & & 
           + \rho \mathrm{g}\ \hat{z} +\nabla\cdot\tau\,,
           \nonumber\\\nonumber\\
      \frac{\partial e}{\partial t} 
          + \nabla\cdot\left[\left(e+\mathrm{p}^{\star}\right)\mathbf{U}
          - \left(\mathbf{U}\cdot\mathbf{B}\right)\mathbf{B}\right] & = &
          + 2 \rho\ \Omega^{2} q x\ \hat{x}\cdot\mathbf{U} \nonumber\\ & &
          + \rho \mathrm{g}\ \hat{z}\cdot\mathbf{U} 
          + \nabla\cdot\tau \mathbf{U} \nonumber\\ & &
          + \nabla\cdot\left[\eta\mathbf{B}\times\left(\nabla\times\mathbf{B}\right)\right]
        \nonumber\\ & & 
          + \nabla\cdot\kappa \nabla {T} 
          - \rho^2 \Lambda\left({T}\right)\nonumber\\ & &
          + \Gamma_{\mathrm{SN}}+\rho\ \Gamma\left(z\right)\,,
          \nonumber\\\nonumber\\
      \frac{\partial\mathbf{B}}{\partial t} 
          - \nabla\times\left(\mathbf{U}\times\mathbf{B} 
          - \eta_m\nabla\times\mathbf{B}\right)&=& 0\,,
          \nonumber\\\nonumber\\
      \frac{\partial e_c}{\partial t} 
          + \nabla\cdot\left(e_c\mathbf{U}+ \mathcal{F}_c\right) &=&
          - \mathrm{p}_c\nabla\cdot\mathbf{U} + \mathcal{Q}_c\,,
\label{model}
\end{eqnarray}
The first four equations in the set represent mass conservation, 
momentum conservation, total energy conservation and induction 
equation respectively, similar to \cite{bendre_15}, and all the 
symbols carry their usual meanings eg. $\rho$, $\mathbf{U}$, $e
$ and $\mathbf{B}$ denoting the density, velocity, total energy 
and magnetic field respectively. The last equation is an addition 
and it describes the time evolution of CR energy density $e_c$. 
The associated notations used therein namely ${\rm p}^{*}$, ${
\rm p}_c$ and $Q_c$ etc. are defined in the following paragraphs. 
Here the propagation of CR energy density is modelled using the field 
aligned diffusion prescription, encapsulated in the anisotropic 
diffusion for CR energy flux $\nabla\cdot\mathcal{F}_c$. We 
follow the non-Fickian prescription similar to \cite{snodin2006} 
to compute this flux term. Specifically we solve the following 
telegraph equation to obtain the evolution of $ \mathcal{ F}_c$ 
simultaneously with 
\eref{model}. 
\begin{align}
    \frac{d \mathcal{F}_{ci}}{d t} = \frac{1}{\tau_c} \left(K_{ij}\nabla_j e_c - \mathcal{F}_{ci} \right)      
\label{flux_non_fick}
\end{align}
The diffusion coefficient $K_{ij}$ is expressed by 
\begin{align}
    K_{ij} = K_{\perp} \delta_{ij} + K_{\parallel}\Hat{\mathbf{B}}_i\Hat{\mathbf{B}}_j   
\end{align}
where $K_{\perp}$ and $K_{\parallel}$ are diffusion coefficients 
in perpendicular and parallel to the direction of local magnetic 
field respectively, and $\Hat{\mathbf{B}}_i$ and $\Hat{\mathbf{B
 }}_j$ are $i$ and $j^{\rm th}$ components of the unit vector in 
the direction of magnetic field $\hat{ \mathbf{B}} =\mathbf{B}/|
\mathbf{B}|$. 

This non-Fickian prescription of field aligned diffusion 
of CR energy (\eref{flux_non_fick}) is preferred instead 
of the standard Fickian one (eg. $\mathcal{F}_{ci}=-K_{i
j} \nabla_{j} e_c$) in order to restrict the propagation 
speed of diffusion to the finite values especially in 
vicinity of magnetic field configuration such as an `x' 
point (eg. $\mathbf{B}=\left(\sin{\left(\pi x/L_x\right)
}, ~ \sin{ \left(-\pi y/L_y\right)},~0\right)$). This is 
achieved by choosing a finite value for the correlation 
time $\tau_c$ in \eref{flux_non_fick}. Solution to \eref{flux_non_fick} approaches to the Fickian diffusion flux, 
as the estimated value of Strouhal number, $S_{t}=\sqrt{K_{
 \parallel} \tau_c }  /h $ (where $h$ is the length scale) 
approaches zero. For the opposite extreme, when $ S_{t} $ 
attains higher values, solution to \eref{flux_non_fick} 
becomes oscillatory. For the models presented in here the value of $S_{t}$ is $\sim 10^{-2}$, and therefore the solution is very similar to the one expected when using the 
standard Fickian prescription for the CR flux 
\citep[see eg.][]{bendrethesis}. In addition to the field 
aligned diffusion we have also incorporated a small 
isotropic diffusion term for CR energy, with a diffusion 
coefficient much smaller than both $K_{\parallel}$ and 
$K_{\perp}$, this is mainly for the numerical reasons. 

Furthermore the term $\nabla\cdot e_c\mathbf{U}$ encapsulates 
the advection of CR energy. An extra pressure term $\mathrm{ 
p}_c$ due to CR energy, back reacts on the flow velocity 
through the Navier-Stokes equation wherein $\mathrm{p}^\star 
$ represents the total pressure, that is thermal pressure, 
magnetic pressure and CR pressure. First two of these pressure 
contributions have already been discussed in \cite{bendre_15} 
and \cite{bendrethesis}, and this extra one is calculated as 
$\mathrm{p}_c = \left(\gamma_c -1\right)e_c$ (where the used 
value of $\gamma_c =14/9$, similar to \cite{ryu2003}). 
Additionally, the term $ \mathrm{p}_c \nabla\cdot\mathbf{U}$ 
models the adiabatic heating effect for CR component. 
Furthermore the term $\mathcal{ Q}_c$ represents the rate at 
which CR energy ($e_c$) is injected through the SN explosions, 
which occur randomly at predefined rates. The fraction of SN 
energy that goes into the ISM as CR and thermal component are 
chosen as an input parameter. 

We also note here that we have not included the effect of CR 
streaming instability that depends upon the local Alfv\'en 
velocity, encapsulating energy transferred from CR to Alfv\'en 
waves \citep[see eg.][]{blandford,kulsrudbook}, although the 
total time for which the model has been run is not sufficient 
for the magnetic field to reach its equipatition values, and 
the magnitude of Alfv\'en velocity is still negligible 
throughout the run time.

\section{Description of Simulated Models}
\label{sec:model_parameters}
We simulate two models in total, the principle difference in 
these, lies in the fraction of SN explosion energy that goes 
into CR and thermal component. To examine the specific effects 
of CR component on the growth of magnetic field, in one of the 
models \verb|CR|, we inject all the SN energy into CR and for 
the model \verb|CR_TH|, we inject the same amount of SN energy 
into a CR component as that in model \verb|CR|, in addition 
thermal energy is also injected. Fraction of SN energy going 
into ISM as a CR and thermal energy ($e_{cr}$ and $e_{th}$) 
for both of these models, is listed in \tref{tab:sneratio}. 
Furthermore the initial conditions for 
various other parameters (such as mid-plane mass density, 
vertical scale height etc.) are set so as to mimic the ISM 
environment in the vicinity of Solar circle in the Milky Way, 
although the azimuthal angular velocity $\Omega$ for both models 
is set to $100$\kms$\kpc^{-1}$, faster 
than that of the Solar circle's neighbourhood. While the rate 
of SN explosion in both models is set to $\sim3\kpc^{-2}\Myr^{
-1}$, about $10\%$ of the SN rate in the Milky Way. 
Initial 
configuration of magnetic field is such that $B_x$ and $B_y$ 
components have strengths of $-10^{-4}$\muG and $ 10^{ -3 }$
\muG respectively at the midplane, with scale-heights of $
\sim 325 \pc$ (equivalent to the initial density scale-heights), also the 
strength of $ B_z$ component is about $\sim 10^{-3}$ \muG, 
throughout the box with vertical flux of about $ 0.0064$
\muG$\kpc^{2}$ threading the $x-y$ plane. 
\begin{table}
\centering
\begin{tabular}{c c c}
\hline
    Model       & $e_{c}$   & $e_{th}$      \\
            & $\erg$       & $\erg$           \\
            \hline
    \verb|CR|   & $10^{50}$   &$0$              \\
    \verb|CR_TH|& $10^{50}$   &$9\times10^{50}$\\
    \hline
\end{tabular}
\caption{Distribution of SN explosion energy}
\label{tab:sneratio}
\end{table}
For the CR diffusion coefficients we choose $K_{\parallel}=3\times 
10^{ 27}\,{\rm cm}^2{\rm s}^{-1}$ and the ratio $K_{\parallel}/K_{
\perp}$ of 100. These are at least an 
order of magnitude smaller than the effective diffusion coefficients 
expected for \muG strength magnetic fields \cite[eg.][]{ryu2003}. This is done mainly to 
constrain the time step and have a sufficiently long simulation in 
realistic times. Moreover these diffusion coefficients are expected to depend on the CR energy itself, roughly as $K_{\parallel
}\propto e_c^{-0.3} $, \citep[see eg.][]{navagabici}, and on the intermittency 
of the magnetic fields \citep[see eg.][]{bera_2017}, both of these effects are 
not incorporated here. This choice of diffusion coefficients likely 
impacts the multiphase morphology of ISM. We therefore categorically 
refrain from analyzing the impact of the inclusion of CR energy on 
the properties of ISM and the volume filling fractions of various ISM 
phases etc. in this work and focus mainly on the growth of magnetic 
field and the properties of dynamo thus realized.

\section{Results}
\label{sec:results}
To analyze the results of these simulations in the context of 
dynamo we first define the average/mean of the flow variables 
${\mathbf{U}}$ and ${\mathbf{B}}$ by integrating them over the 
horizontal planes and express the mean quantities as the 
functions of $z$;
\begin{align}
\mean{\mathbf{U}}\left(z,t\right) &=\frac{1}{L_x\,L_y} \iint\mathbf{U}\,{\rm d}x{\rm d}y,\nonumber\\
\mean{\mathbf{B}}\left(z,t\right) &=\frac{1}{L_x\,L_y} \iint\mathbf{B}\,{\rm d}x{\rm d}y.
\label{mean}
\end{align}
This definition allows one to express the local velocity and 
magnetic fields as the sums of their respective mean and 
fluctuating components, $\mathbf{U} = \mean{\mathbf{U}} + 
\mathbf{u}$ and $\mathbf{B} = \mean{\mathbf{B}} + \mathbf{b}$. 

\subsection{General Evolution}
\begin{figure}
    \centering
    \includegraphics[width=1\linewidth,keepaspectratio]{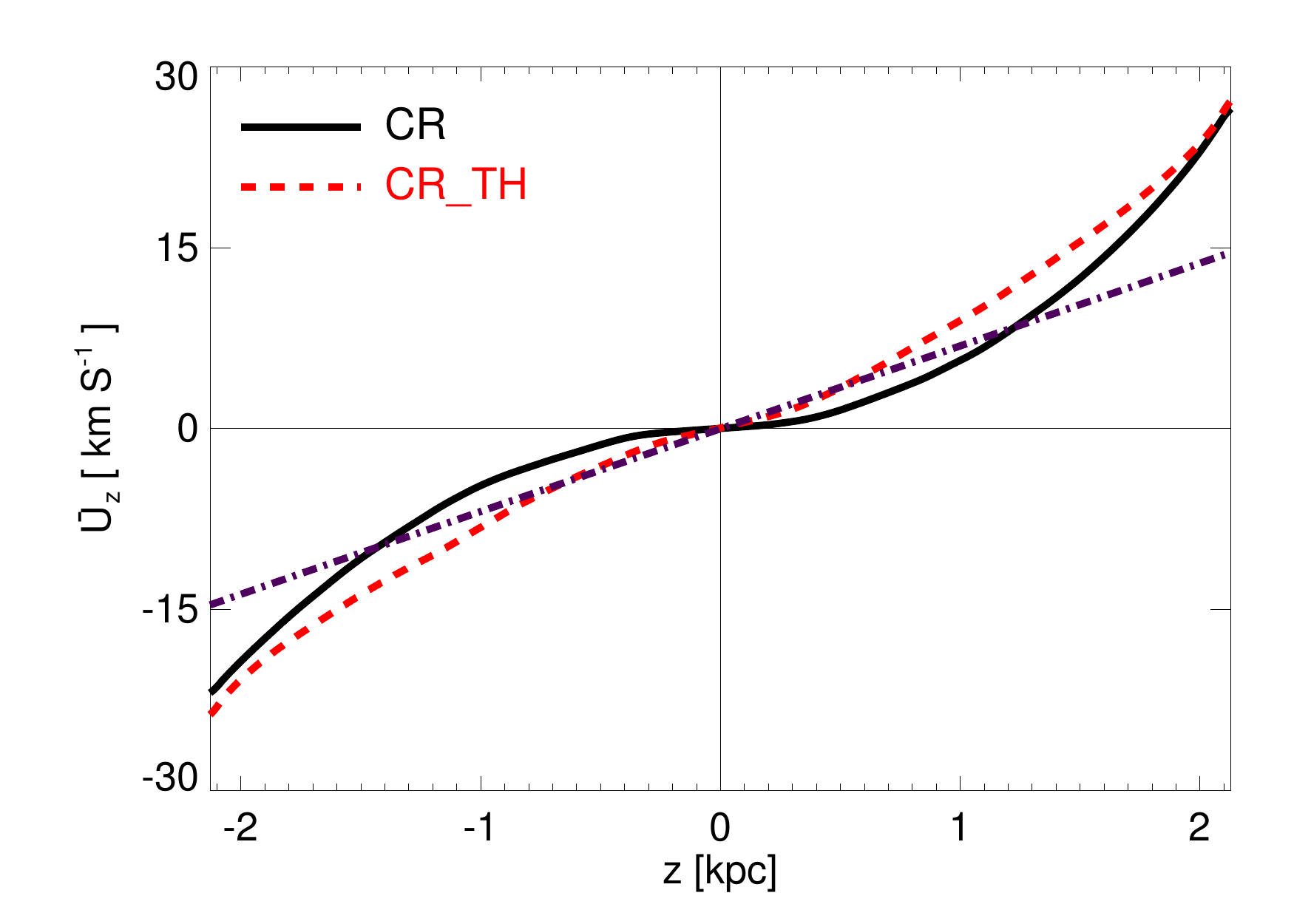}
\cprotect\caption{Black solid line shows the stationary vertical 
    profile of outward wind, $\overline{U}_z$ in model \verb|CR|, 
    while red-dashed line represents the same for model 
    \verb|CR_TH|. The dot-dashed line in color blue represents 
    the expected vertical profile of wind for the used value of 
    SN rate here. The SN rate scaling derived from the trend seen 
    in our previous simulations without CR component.}
    \label{fig:uz}
\end{figure}

Within first $\sim 50 - 100\Myr$ kinetic, thermal and CR energy 
in both models reach a quasi-stationary state, with CR energy 
being the largest contribution to overall energy budget. This is 
presumably due to an almost order of magnitude weaker values of 
diffusion coefficients used in our simulations. The magnitude CR 
energy in this quasi-stationary state is almost 2-3 times higher 
in \verb|CR_TH|, a model that includes the CR and thermal energy 
as an input from the SN explosions, than in the model \verb|CR|.
This could be attributed to the adiabatic rise (expressed by the
term $p_c\nabla \cdot \mathbf{U}$ in the CR propagation equation
), due to higher local velocities in model \verb|CR_TH|. This 
quasi-stationary state is also associated with emergent steady
vertical profiles of mean outward wind velocity $\overline{U}_z$ 
(averaged over $x-y$ plane), these profiles are comparatively 
flatter in the inner part of the disc $|z|<1\kpc$ than in the 
outer parts where these increase quadratically, at $z=2\kpc$, 
magnitude of $ \overline{U}_z $ is about $24$\kms. This increase is more prominent for both models with 
CR element than that for those which include purely thermally 
driven SN explosions. From our previous simulations we were able 
to derive a dependence of the magnitude of $\overline{U}_z$ on 
the rate of SN explosions, the trend goes roughly as 
$|\overline{U}_z|\sim \sigma^{0.4}$, where the $\sigma$ stands 
for the normalized SN explosions' rate \cite[similar to ][ referaces therein]{gressel_2012}. Magnitudes of 
$\overline{U}_z$ at the outer boundaries in the CR simulations is almost twice as high 
compared to its value expected from this SN rate scaling, a 
difference that could also be attributed to a different vertical 
hydrodynamic equilibrium due to additional CR pressure. For the 
model \verb|CR_TH| on the other hand the inner profiles of $\mean{
U}_z$ (between approximately $|z| < 1\kpc$) tend to be similar 
to the once expected for this SN rate. These vertical profiles 
of wind are shown in \fref{fig:uz} for both models.

{}

\subsection{Evolution of Magnetic Energy}
\begin{figure*}
    \centering
\includegraphics[width=0.9\linewidth,keepaspectratio]{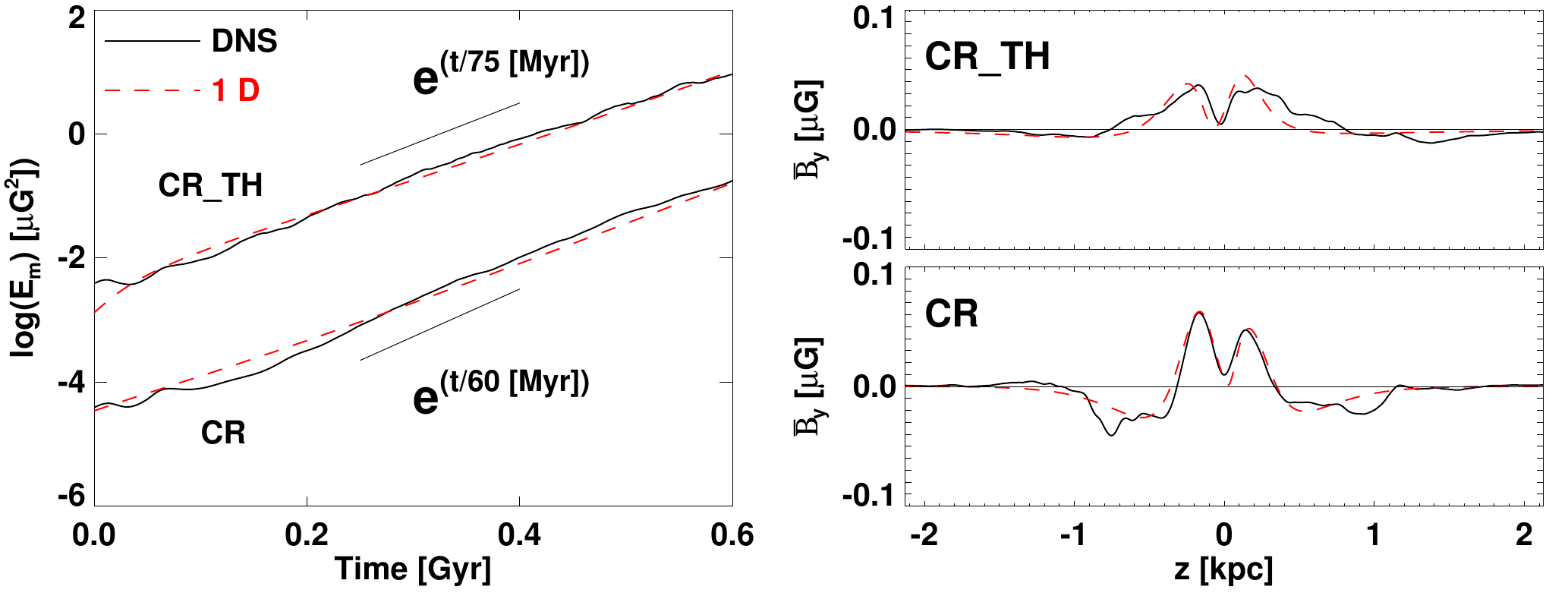}
\cprotect\caption{{\it Left Panel} Solid-black lines show the 
amplification of mean magnetic energy in models \verb|CR| and 
\verb|CR_TH| shown in logarithemic units. We have multiplied 
the magnetic energy in model \verb|CR_TH| by a constant factor 
of 2, to avoid the overlap. Overplotted in red dashed lines 
are the time evolution plots of magnetic energy in respective 
models calculated using 1-D dynamo equations. Note the overlap 
of both curves, indicating the a qualitative similarity in the 
evolution magnetic energy. {\it Right Panel} Black-solid lines 
show the vertical profiles of the azimuthal components of 
magnetic field after $600\Myr$ of evolution for model \verb|CR| 
(bottom panel), and for the model \verb|CR_TH| (top panel).
Red dashed lines plotted in both panels also show the vertical 
profiles of $\overline{B}_y$ after $600 \Myr$ in both models but 
calculated using the 1-D dynamo equations. Note the approximate 
similarity of these profiles.}
    \label{fig:by_ene}
\end{figure*}

After the initial mixing phase of $\sim50\Myr$, magnetic energy $E_m$ in both models amplifies exponentially at e-folding times 
of approximately 60 and $75\Myr$ for models \verb|CR| and \verb|CR_TH|, 
respectively, this time evolution is shown in the left hand panel 
of \fref{fig:by_ene} with black solid lines. Overall these growth 
rates are much faster than that for the purely thermal SN models 
discussed in \cite{bendre_15} where the e-folding time was found 
to be $\sim100\Myr$ irrespective of SN rate. This hints at the 
enhancement of growth rate of dynamo with inclusion of the CR 
component. Vertical profiles of mean magnetic field components 
$\overline{{B}}_x$ and $\overline{{B}}_y$ in both 
models amplify exponentially and within $ \sim 100\Myr$ attain a 
vertically symmetric profile. For the model \verb|CR_TH| these 
vertical profiles are double peaked, with their maximums located 
at $\sim\pm 200\pc$, and the scale height of approximately $\sim 
1 \kpc$. These are also superimposed with comparatively smaller 
peaks at $\sim\pm1.2\kpc$ with opposite signs than the inner peaks 
located at $\sim\pm200\pc$. While for the purely CR model 
\verb|CR| these vertical profiles are similar to those in 
\verb|CR_TH| except the negative peaks are located at $\sim 0.8 
\kpc$ and are much more pronounced. This is seen clearly with black 
solid lines in the right hand panels of \fref{fig:by_ene}, where 
we show the vertical profiles of $\mean{B}_y$ after $0.6\Gyr$ for 
both models. Overall the mean field profiles are wider in the model 
\verb|CR_TH| perhaps due to additional advection present due to 
thermal energy. However, as a consequence of the lack of advection, 
in the pure CR model \verb|CR| mean magnetic field tend to stay 
longer in the inner dynamo active region. This leads to a slightly 
faster growth rate of mean-fields on the model \verb|CR| as shown 
in the right hand panel of \fref{fig:by_ene}.

This peculiar shape of $\overline{{B}}_x$ and $\overline{
{B}}_y$ profiles is markedly different from the ones seen 
in purely thermal SN models, where a vertically symmetric profile 
with a single peak located at $z=0$ was consistently obtained 
regardless of the SN explosions rate. Although a similar vertical 
profiles of mean field have been shown to evolve in another 
similar setup \citep[see Fig. 3. of][]{hanasz2004}.

\subsection{Mean Field Formulation}
In these simulations both $\overline{\mathbf{B}}$ and $\mathbf{
b} $ amplify exponentially at the e-folding time of $150$ and $
120\Myr$ approximately, for models \verb|CR_TH| and \verb|CR| respectively. In 
order to understand this amplification and implications of CR 
component for it, we use the standard mean-field dynamo 
formulation \citep{krause_radler_1980}, while relying on the definition of 
mean given in \eref{mean}. In the dynamo framework, one seeks 
to understand the growth of mean magnetic field $\mean{\mathbf{
B}}$ for a given background flow $\overline{\mathbf{U}}$ 
\citep[eg.][]{Mof78}. By substituting the magnetic and velocity 
field as a sum of mean and turbulent components in the induction 
equation, one obtains the induction equation for the mean 
field, which can be written as 
\begin{align}
\frac{\partial \overline{\mathbf{B}}}{\partial t} = \nabla \times \left( \overline{\mathbf{U}}\times\overline{\mathbf{B}}-\eta_m\nabla\times\overline{\mathbf{B}} +\overline{\emf}\right)
\label{dynamo_eq1}
\end{align}
This is similar to the evolution equation of total magnetic field 
$ \mathbf{B} $ eg. \eref{model}, except for an extra EMF term 
$\overline{\emf}=\overline{\mathbf{u}\times\mathbf{b}}$, which 
serves as a main driver for the amplification process. By employing 
the widely used Second Order Correlation Approximation (SOCA) 
\citep[eg.][]{radler2014}, components of EMF are modeled as linear 
functional of mean magnetic field and its first derivatives,
\begin{align}
\overline{\emf}_i = \alpha_{ij}\overline{B}_j - \eta_{ij}\left(\nabla\times\overline{\mathbf{B}}\right)_j,
\label{emf}
\end{align}
where the tensorial quantities $\alpha_{ij}$ and $\eta_{ij}$ 
represent the dynamo coefficients that depend on the properties 
of background turbulence. Here the diagonal components of 
$\alpha_{ij}$ encapsulate the classical `alpha' effect 
originating from the net helicity of the turbulent motions, 
while the off-diagonal ones represent the `turbulent pumping effect' arising from the gradient of turbulent 
intensity. On the other hand $\eta_{ij}$ tensor's diagonal 
components represent the turbulent magnetic-diffusivity effect, 
and off-diagonal ones represent the R\"adler effect 
\cite{radler69} \citep[see also][ for more discussion]{BS05}.
%and the shear-current effect \cite{rogachevskii} \citep[see %also][]{BS05,axel_radler}. 
These 
interpretations become clear when 
\eref{dynamo_eq1} is expressed in its component form as, 
\begin{align}
\frac{\partial\overline{{B}}_x}{\partial t}& =
\frac{\partial}{\partial z} \Bigg[
-\left(\overline{U}_z +\alpha_{yx}\right)   \overline{{B}}_x
-\alpha_{yy}    \,                          \overline{{B}}_y
+\eta_{yy}     \, \frac{\partial\overline{{B}}_x}{\partial z}              
-\eta_{yx}     \, \frac{\partial\overline{{B}}_y}{\partial z}                         
\Bigg]\nonumber\\ \nonumber\\
\frac{\partial\overline{{B}}_y}{\partial t} &=
\frac{\partial}{\partial z} \Bigg[
-\left(\overline{U}_z -\alpha_{xy}\right)   \overline{{B}}_y
+\alpha_{xx}     \,                         \overline{{B}}_x
+\eta_{xx}       \, \frac{\partial\overline{{B}}_y}{\partial z}                   
-\eta_{xy}       \, \frac{\partial\overline{{B}}_x}{\partial z}
\Bigg]\nonumber\\&+q\,\Omega\,\overline{{B}}_x\nonumber\\\nonumber\\
\frac{\partial\overline{{B}}_z}{\partial t} &= 0.
\label{dynamo_eq2}
\end{align}
Here the $\alpha_{xy}$ and $\alpha_{yx}$ appear clearly in 
the advection term involving $\mean{U}_z$ while the diagonal 
$\alpha_{ij}$ and $\eta_{ij}$ appear as the source and diffusive 
terms respectively. 

Advantage of this type of analysis is the ability to self 
consistently probe which aspect of turbulent motions are 
contributing to the amplification process, and to understand 
the manner in which the additional component of CR is affecting 
the background turbulence and in turn the evolution of mean field. 
In the following subsection we qualitatively discuss the 
computed profiles of dynamo coefficients and effects of CR 
component upon them.

\subsection{Dynamo Coefficients}
\begin{figure*}
    \centering
    a)
    \includegraphics[width=0.46\linewidth,keepaspectratio]{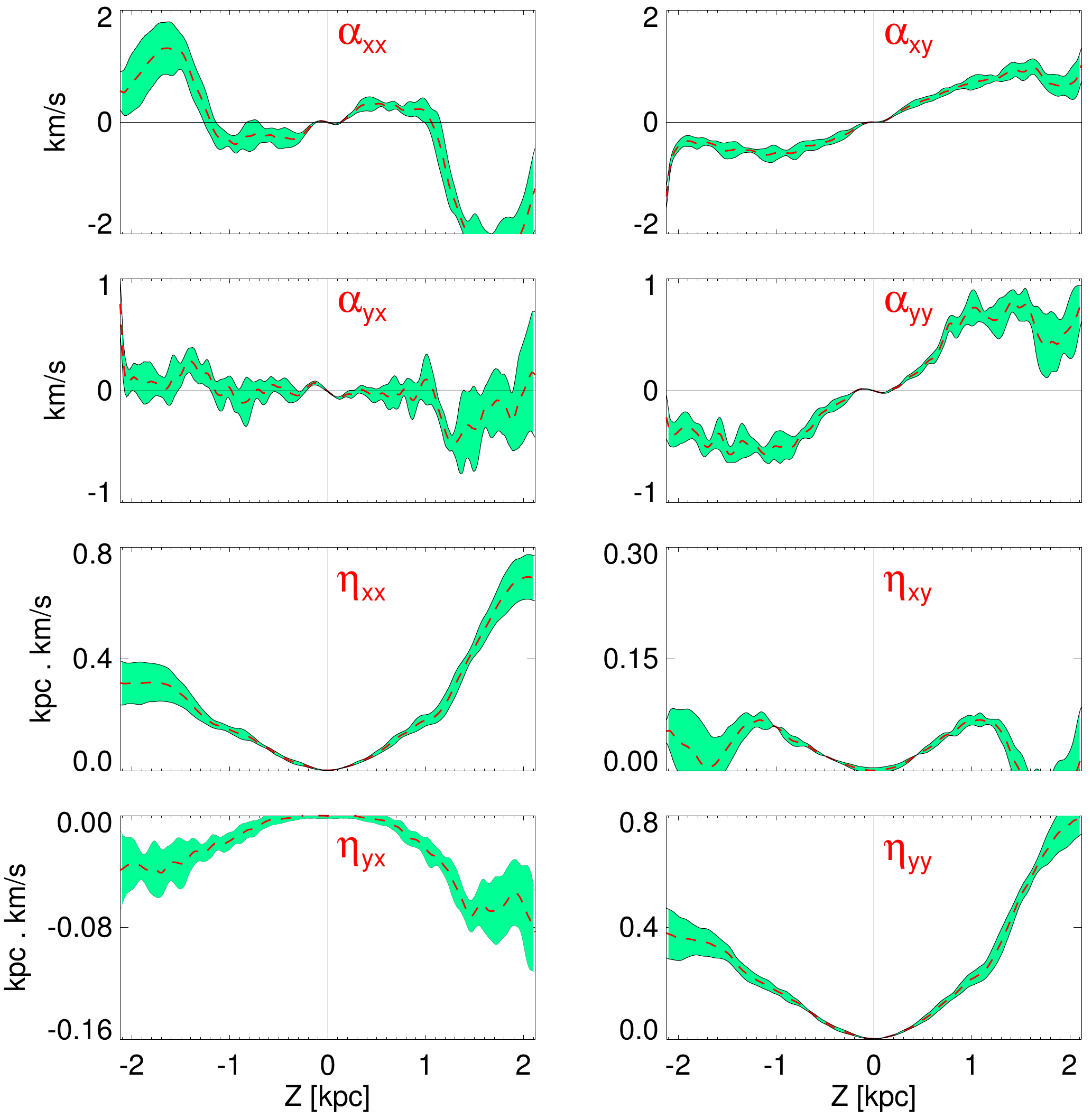}
    \hspace{0.5cm}
    b)
    \includegraphics[width=0.46\linewidth,keepaspectratio]{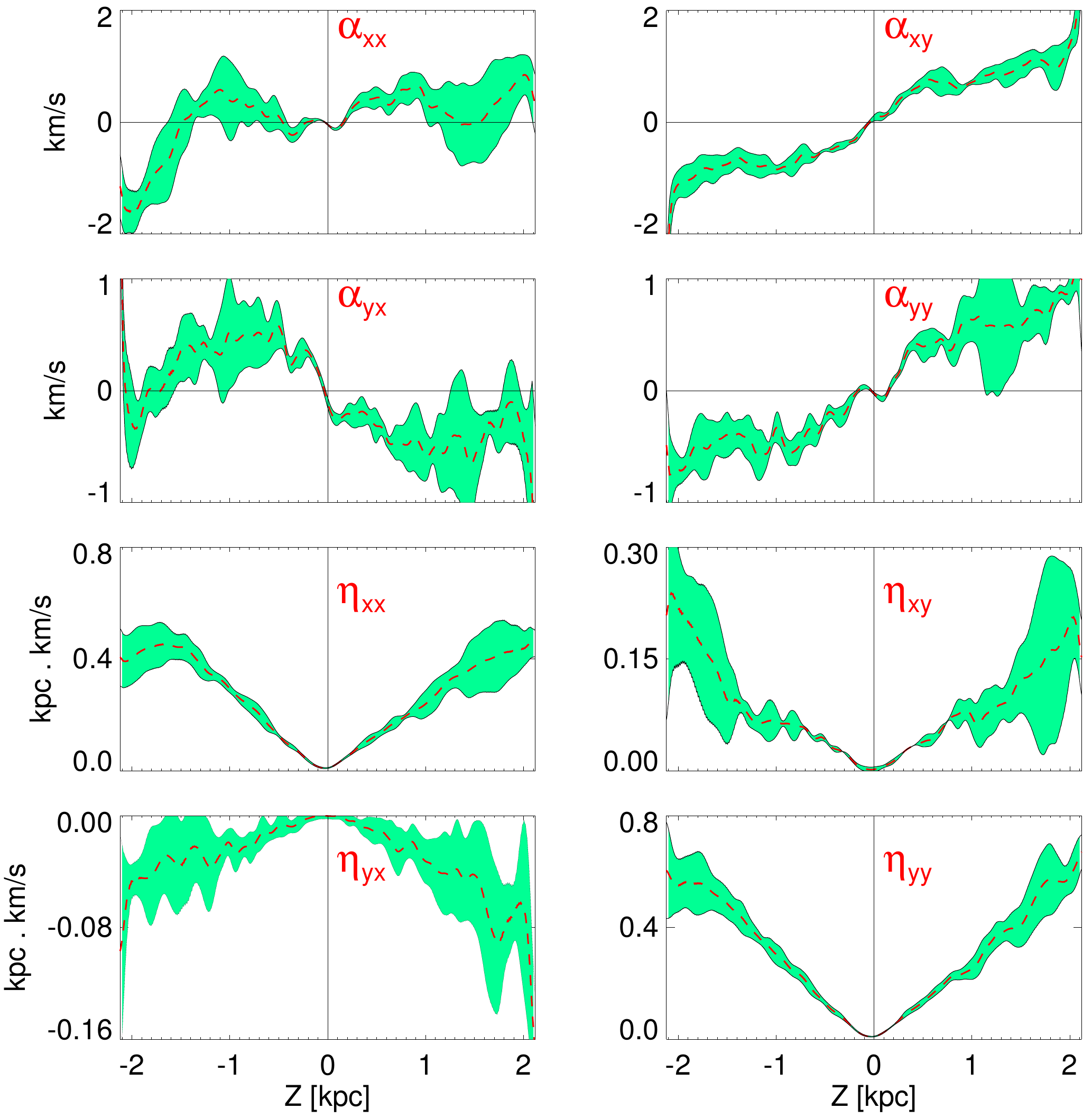}
\cprotect\caption{{\it Panel} (a) Red dashed lines indicate the 
    vertical profiles of dynamo coefficients for model \verb|CR|, 
    along with corresponding $ 1 - \sigma$ error estimates shown 
    in green shaded regions. {\it Panel} (b) similar to {\it 
    Panel} (a) but for model \verb|CR_TH|.}
    \label{fig:alpha_eta_profiles}
\end{figure*}

We use the standard test-fields method to extract the 
components of dynamo tensors $\alpha_{ij}$ and $\eta_{
ij}$. We have already used this method for a similar
setup, in our previous analysis \cite{bendre_15}, to 
calculate the dynamo coefficients in the models with 
purely thermal SN explosions. More details about the 
test-fields methods implementation and caveats are 
discussed in \cite[][]{brandenburg2009,brandenburg2018,nonlocality} etc. 
Profiles of all such coefficients as functions of $z$, 
we have thus computed, are shown in the left and right 
hand panels of \fref{fig:alpha_eta_profiles} for model 
\verb|CR| and \verb|CR_TH| respectively, along with 
their corresponding $ 1 - \sigma $ error estimates 
shown in green color shade. It is manifestly clear from 
the figures that the diagonal components of $\alpha_{ij}
$ are almost 40 - 50\% weaker in model \verb|CR| than in 
model \verb|CR_TH|. Quantitatively speaking at $z=1\kpc$ $\alpha_{xx} $ is approximately $0.3\pm0.08$\kms in model \verb|CR| while it is $\sim 0.6\pm0.15$\kms in model \verb|CR_TH|, although it is much noisier than other dynamo coefficients. Whereas $\alpha_{yy}$ component is $0.5\pm0.15$\kms and $0.8\pm0.20$\kms in model \verb|CR| 
and \verb|CR_TH| respectively. Although for both models, 
the dynamo coefficients are still almost 4-8 times 
smaller than their expected values estimated from the 
trends with respect to SN rate, observed in our previous
simulations without CR, highlighting the difference 
between CR and thermal SN turbulence. 

Another aspect in which the CR models differ from 
the previous models without the CR component is the 
one of off-diagonal components of both $\alpha_{ij}
$ and $\eta_{ij}$ tensors. In particular, the 
antisymmetric contribution from the $\alpha$ tensor 
tends to be negligible in the \verb|CR| model, in a 
sense that magnitude-wise $\alpha_{yx}$ component is 
negligible compared to $\alpha_{xy}$ implicating an 
absence of a systematic turbulent pumping effect 
$\gamma$. This appears in the mean induction equation 
as $\gamma \times\overline{\mathbf{B}}$, acting as an 
advective term, roughly in the direction of the 
gradient of turbulent intensity. For the model 
\verb|CR_TH| on the other hand although the $\alpha_{
yx}$ is non-negligible, there is only an approximate 
antisymmetric off-diagonal part, and therefore there 
exists a systematic pumping. This is in contrast 
with the outcomes of our previous simulations of SN 
driven turbulence without the CR component. In those 
simulations the turbulence driven by the thermal SN 
explosions led to an anti-symmetrical off diagonal 
component of $\alpha_{ij}$ tensor (such that $\alpha_{
xy} = - \alpha_{yx}$) or a turbulent pumping term 
that acted against $\mean{\mathbf{U}}_z$ preventing 
the loss of large scale helicity. This subsequently 
led to the amplification of mean-field. Inclusion of 
CR component seems to introduce the seen anisotropy 
in the pumping of $x$ and $y$ components of mean 
field, whereas with only the thermal energy injected 
through SN explosions, the mean magnetic field is 
isotropically pumped in a sense that $x$ and $y  $ 
components are transported via the $\alpha_{yx}$ and
$\alpha_{xy}$ respectively, to a same extent. This
anisotropy is possibly a result of the difference 
between the propagation mechanisms of CR and thermal
energy, while the former propagates 
preferentially along the magnetic field lines as 
prescribed by the field dependent diffusion of $e_c$, 
the later has no such systematic systematic dependence 
on the direction. The negligible contribution of 
$\alpha_{yx}$ in model \verb|CR| also hinders the 
growth of $x$ component of mean field compared to 
its $y$ component, as the $\overline{ U}_z$ overtakes 
(see eg. \eref{dynamo_eq2}, where we write the 
\eref{dynamo_eq1} in its component form). This effect 
is less severe in the model \verb|CR_TH| because of 
the non-zero $\alpha_{yx}$ acting against the vertical 
wind. Such anisotropy has implications for the pitch 
angles of magnetic fields, that is the angle made by 
magnetic field line with respect to the azimuthal 
direction (approximately $\tan^{-1}( \overline{B }_x/
\overline{B}_y)$). As a consequence, in model \verb|CR| 
the calculated values of pitch angles are much smaller 
compared to that in model \verb|CR_TH|. This is markedly 
different than our previous results without the CR where 
with the increasing SN rate, outward wind helped 
saturate the mean magnetic fields to the strengths 
insufficient to quench the dynamo coefficients and 
therefore retained the pitch angles even in the saturated
phase of magnetic fields. 

To further cite the differences in the dynamo coefficients, we 
refer to the components of $\eta_{ij}$ tensor. It appears that 
shapes and trends of the vertical profiles are qualitatively 
similar in both models, and for the diagonal components even 
their strengths are similar. The off-diagonal components however 
are twice as strong in model \verb|CR_TH| than in model \verb|CR|. 
Comparative contribution of the diagonal to off-diagonal elements 
is also much higher than in our previous simulations described in 
\cite{bendre_15}. The ratios of diagonal to off-diagonal 
components were found to be at most 10, in our previous simulations 
irrespective of the SN rates, in fact within the standard $1-\sigma
$ error intervals both $\eta_{xy}$ and $\eta_{yx}$ were consistent 
with zero. However for both models with the CR this ratio is about 
2 to 5. The values diagonal $\eta_{ij}$s calculated for both 
\verb|CR| and \verb|CR_TH| using the test-fields method are almost 
3-6 times weaker than their expected magnitude extrapolated from the 
dependence of $\eta_{ij}$ components on the rate of SN explosions 
derived in \cite{bendre_15} for the models without the CR component. 
However both $\eta_{xy}$ and $\eta_{yx}$ turn out to have the 
expected magnitudes; more or less from the same trend, although it 
should be noted that in the said previous simulations the 
off-diagonal $\eta$ coefficients were much noisier. Consequently the 
estimated value of dynamo number is $\sim190$ for both CR models as 
opposed to $\sim 110$ for the models without the CR component. 

Remarkably, and contrary to the models without the CR, signs of 
the $\eta_{yx}$ and $\eta_{xy}$ components very consistently turn 
out to be negative and positive respectively, in both models with 
the CR component, which in combination with the differential 
rotation and shear is shown to lead to the R\"adler effect and 
amplify the mean fields without any systematic $\alpha$ effect. 
For both models with CR, this effect enhances the growth rate of 
dynamo by $\sim 17$ to $30\%$, as discussed in the following 
section.

\subsection{Evolution of Mean-Field}
To understand the entire evolution of mean magnetic field in both 
models we simultaneously solve \eref{dynamo_eq2} for $\mean{B}_x$
and $\mean{B}_y$.
%Eq. \ref{dynamo_eq1}. By noting that with a 
%used definition of mean only the $z$ component of gradient stays 
%non-zero, we first write the Eq. \ref{dynamo_eq1} in its component 
%form, as; 
%\begin{align}
%\frac{\partial\overline{{B}}_x}{\partial t}& =
%\frac{\partial}{\partial z} \Bigg[
%-\left(\overline{U}_z +\alpha_{yx}\right)   \overline{{B}}_x
%-\alpha_{yy}    \,                          \overline{{B}}_y
%+\eta_{yy}     \, \frac{\partial\overline{{B}}_x}{\partial z}              
%-\eta_{yx}     \, \frac{\partial\overline{{B}}_y}{\partial z}                         
%\Bigg]\nonumber\\ \nonumber\\
%\frac{\partial\overline{{B}}_y}{\partial t} &=
%\frac{\partial}{\partial z} \Bigg[
%-\left(\overline{U}_z -\alpha_{xy}\right)   \overline{{B}}_y
%+\alpha_{xx}     \,                         \overline{{B}}_x
%+\eta_{xx}       \, \frac{\partial\overline{{B}}_y}{\partial z}                   
%-\eta_{xy}       \, \frac{\partial\overline{{B}}_x}{\partial z}
%\Bigg]\nonumber\\&+q\,\Omega\,\overline{{B}}_x\nonumber\\\nonumber\\
%\frac{\partial\overline{{B}}_z}{\partial t} &= 0.
%\label{dynamo_eq2}
%\end{align}
The component $\overline{B}_z$ does not evolve in time as a 
consequence of the solenoidity constraint along with the boundary 
conditions for magnetic field. As an input we choose the vertical 
profiles of dynamo coefficients calculated using the test-fields 
method discussed in the previous subsection, and the profile of 
vertical wind also from the direct simulations. With this setup 
we solve 
\eref{dynamo_eq2} using an algorithm based on finite difference 
method over a staggered grid of size $N_z = 512$. For boundary 
conditions we use the continuous gradient condition equivalent to 
the direct simulations. This is very similar to our previous analysis 
without the CR component, effect of this additional CR element is 
captured wholly in the dynamo coefficients and the vertical wind 
profiles. 

\begin{figure*}
    \centering
    \includegraphics[width=0.8\linewidth,keepaspectratio]{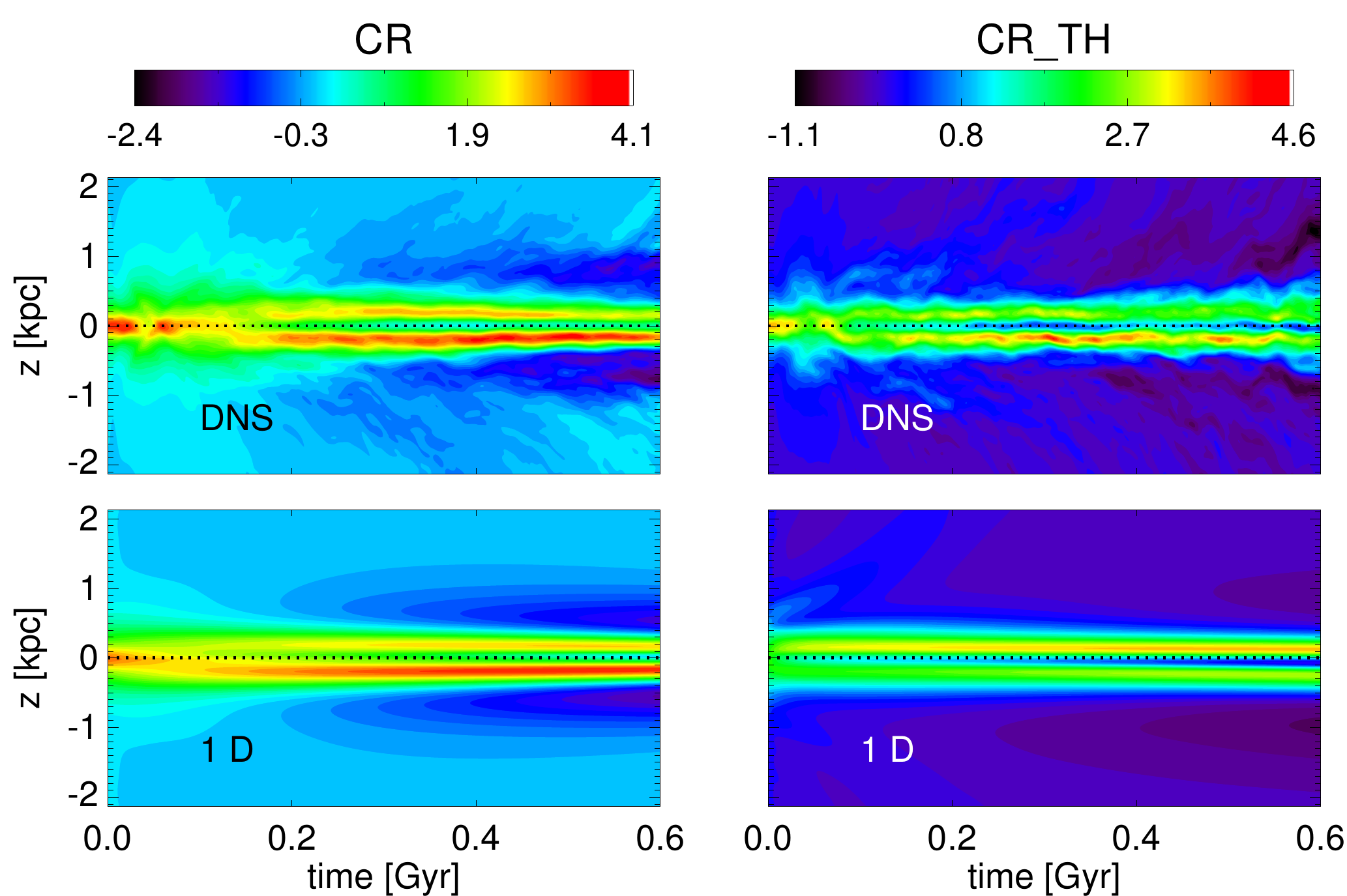}
\cprotect\caption{{\it Left column:} shows the time evolution of 
    the vertical profiles of the $y$ component of mean magnetic 
    field seen in the direct simulations of model \verb|CR| (top
    panel) while the bottom panel shows the same for the 
    corresponding 1D dynamo model using the dynamo coefficients 
    obtained for the model \verb|CR| using the test-fields method.
    Colour code used for producing these contours is defined in 
    the colourbar at the top, however we have scaled it temporally; 
    by a factor of $\exp{(t/120 {\rm \,Myr})}$ to compensate for 
    the exponential growth of the mean magnetic field. {\it Right
    Column} indicated the same as the left column but for the model
    \verb|CR_TH|. Note the qualitative similarity in the evolution 
    of magnetic field in both panels.}
    \label{fig:b_contour_cr}
\end{figure*}

It turns out in this analysis that the evolution of mean field 
predicted using \eref{dynamo_eq2} and dynamo coefficients is 
largely consistent with what obtains in the direct simulations, 
for both models. Which seem to show that the dynamo coefficients 
calculated in the previous subsection do capture the underlying 
dynamics of the turbulence, and the differences marked in the  various dynamo coefficients as discussed 
above also encapsulate the actual impact of CR component on the 
evolution of mean magnetic field. In the left and right hand 
columns of \fref{fig:b_contour_cr} we compare the contours of 
time evolution of the vertical profiles of the $y$ component of 
mean field for models \verb|CR| and \verb|CR_TH| respectively, 
calculated from the DNS and 1-D dynamo models. Note the 
remarkable similarity in the coloured contours shown in top and 
bottom panels which use the same color codes. Furthermore the 
time evolution of mean magnetic energy is also comparable in both 
DNS and 1-D simulations, eg. in the left hand panel of 
\fref{fig:by_ene}, black solid lines (showing the evolution 
of mean magnetic energy in DNS) are comparable to the red-dashed 
ones showing the evolution magnetic energy in 1-D dynamo simulations. 
Also in the right hand panels we compare the vertical profiles of 
azimuthal mean field component at $t=600\Myr$, black solid lines 
show the results from DNS and red-dashed lines are the ones from 
1-D dynamo simulations. There is an overall similarity 
in the evolution of mean field in both types of simulations and it 
can therefore be safely argued that the computed dynamo coefficients 
effectively characterize the turbulence that drives the dynamo. 

%\begin{figure*}
%    \centering
%    \includegraphics[width=0.9\linewidth,keepaspectratio]{fig/by_cont_cr_th.pdf}
%\cprotect\caption{Same as Fig. \ref{fig:b_contour_cr} but for the 
%    model \verb|CR_TH|.}
%    \label{fig:b_contour_crth}
%\end{figure*}

\begin{figure*}
    \centering
    \includegraphics[width=0.48\linewidth,keepaspectratio]{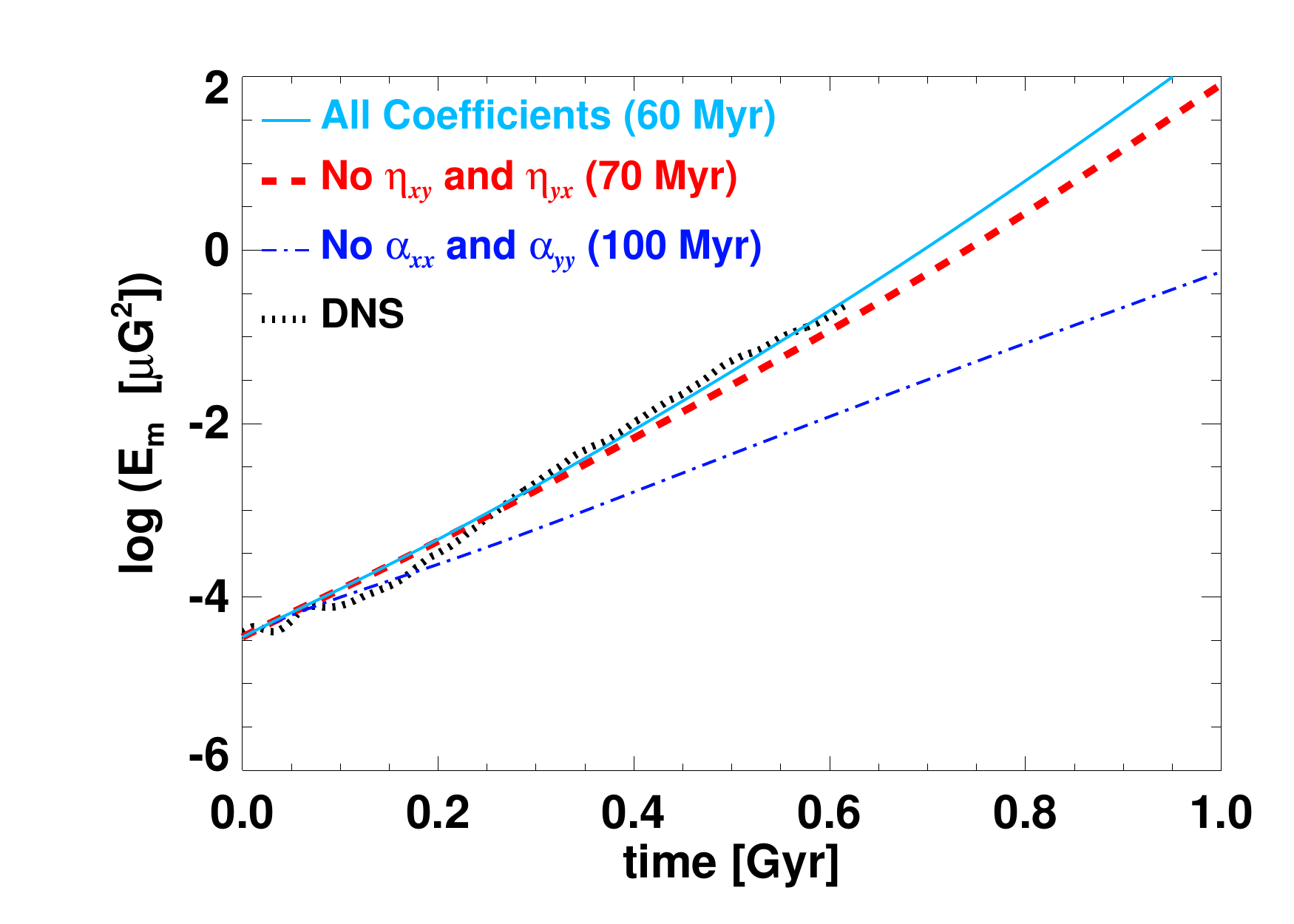}
    \includegraphics[width=0.48\linewidth,keepaspectratio]{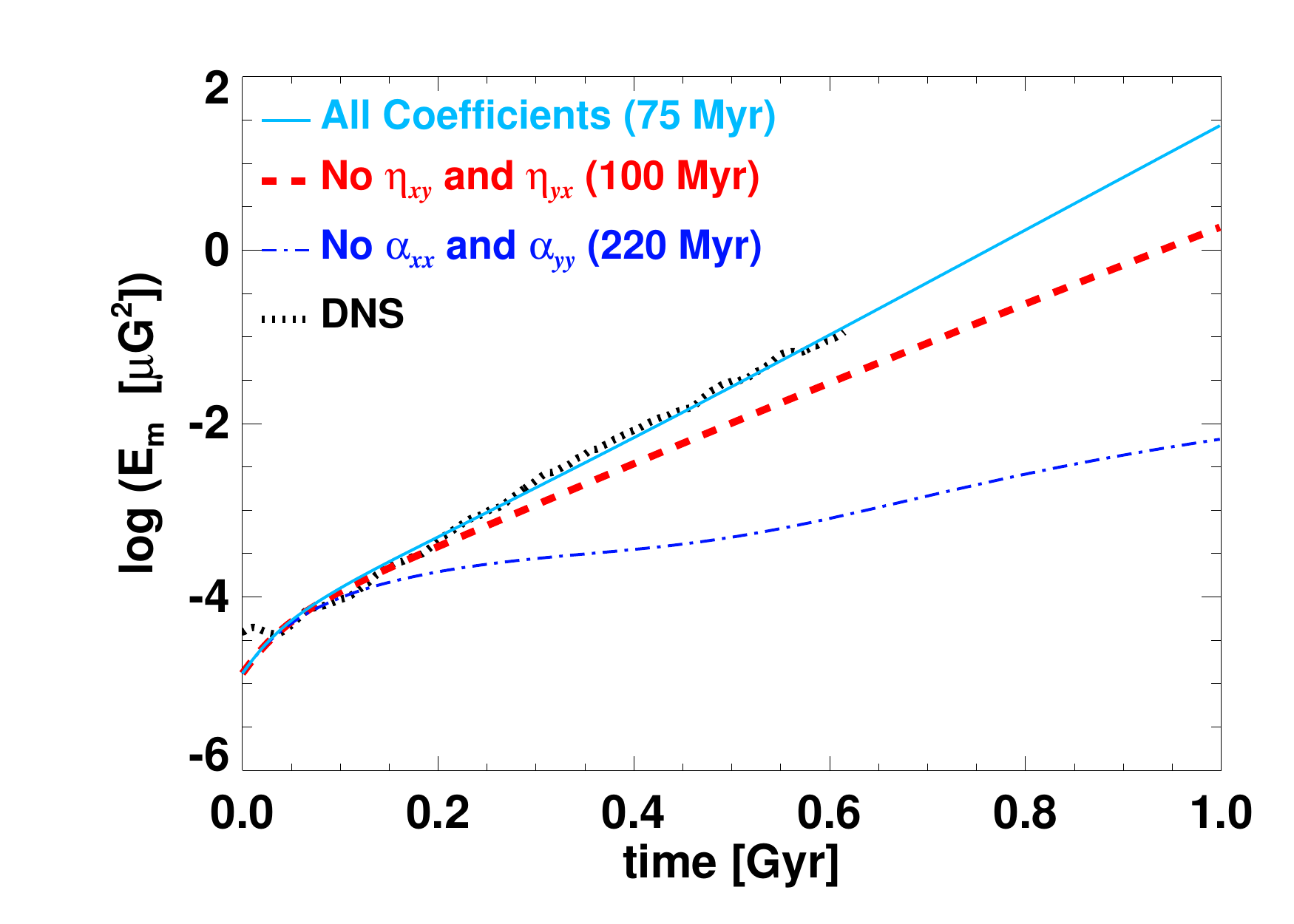}
    \cprotect\caption{{\it Left Panel} shows the 
    time evolution of mean magnetic energy in model \verb|CR|. 
    Black dotted line shows the outcome of DNS while the 
    light blue solid line shows the same for corresponding 
    1 D dynamo model. Red dashed line shows the evolution 
    of mean magnetic energy in a dynamo model where the 
    off-diagonal elements of $\eta$ tensors have been set 
    to zero. The blue dot-dashed line shows a complementary
    case where the diagonal $\alpha$ elements are turned off and the full $\eta$ tensor is used, 
    highlighting that R\"adler effect independently leads 
    to a growing mean field. Approximate e-folding times of
    magnetic energy in each of the cases are written in
    brackets. {\it Right Panel} is same as the Left panel 
    but for the model \verb|CR_TH|. Note that without the
    $\eta_{xy}$ and $\eta_{yx}$ terms the growth of dynamo 
    is hindered noticeably.}
    \label{fig:shear_current}
\end{figure*}

Additionally, to identify the impact of non-negligible 
off-diagonal $\eta$ components on the evolution, we set the 
corresponding terms in \eref{dynamo_eq2} to zero and run the 
1-D simulations with rest of the dynamo coefficients for both 
CR models. Outcomes of these simulations reveal that the 
slightly fastened growth rate of dynamo in \verb|CR| and 
\verb|CR_TH| is indeed a result of R\"adler effect. This 
can be seen from \fref{fig:shear_current} where in left and 
right hand panels we compare the evolution of mean magnetic 
energy 
in models \verb|CR| and \verb|CR_TH| respectively. With black 
dotted lines we show the mean magnetic energy from DNS, 
along with its counterpart from 1-D dynamo simulations in 
light-blue solid lines while the red dashed lines show the 
evolution of the same 1-D dynamo without the off-diagonal 
$\eta_{ij}$ components. We point out here a clear hindrance 
in the growth of the dynamo when $\eta_{xy}$ and $\eta_{yx}$ 
are absent in the simulations. For the model \verb|CR| the 
absence of off-diagonal terms translates to almost a $17\%$ 
decline in the growth rate of mean magnetic energy (e-folding 
time of $60\Myr$ for the model including $\eta_{xy}$ and 
$\eta_{yx}$ as opposed $70\Myr$ without them). Similarly 
for the model \verb|CR_TH| the e-folding time increases from 
$75\Myr$ to $100\Myr$ once the off-diagonal $\eta$ components
are switched-off as can be clearly seen in right panel. 
Interestingly, even in the absence of systematic $\alpha$
effect, this combination of off-diagonal $\eta$ components 
and the differential shear is sufficient for the exponential
growth of dynamo in both models, as shown by the blue
dot-dashed lines. This points towards the R\"adler effect 
dynamo associated with CR turbulence.

Furthermore to investigate the vertical profiles of 
$\overline{U}_z$ that are dissimilar to the ones seen in 
the models without the CR, as a potential reason for the 
enhanced dynamo action, we do a similar exercise. We set 
the wind term in \eref{dynamo_eq2} to zero and perform 
1-D dynamo simulations again. This however reveals that 
the growth rate of dynamo is {\it not} significantly 
impacted by the wind profiles, the difference however, 
is that without the outward wind mean magnetic field is 
concentrated mostly in the inner disc part $\left(-1\kpc 
< z < 1\kpc\right)$ although with a smaller dynamo 
time-period.

\section{Conclusions}
\label{sec:conclusions}
Main effects of CR on the ISM turbulence and mean field 
dynamo seen in our simulations are as follows,

\begin{enumerate}
\item Faster increasing velocities of the outward wind with height.
\item Anisotropy of turbulence transport, which gets reflected in 
the vertical profiles of off-diagonal $\alpha$ coefficients.
\item Appearance of a R\"adler term, which slightly increases 
the growth rate of mean magnetic field.
\item Slightly enhanced growth of the dynamo with inclusion of CR 
term.
\end{enumerate}

\section{Summary}
\label{sec:summary}
We have examined here the impact of the field aligned diffusion of 
CR energy injected through the SN explosions on the evolution of 
galactic dynamo. We have done so by comparing the evolution of 
magnetic fields in the model where only thermal energy was injected 
by SN explosions, to a model that treats the SN explosions as 
localised injection of both thermal energy and the CR energy. We 
treat the CR energy as a fluid diffusing in the direction of local 
magnetic field. We have also compared the general evolution of both 
models to our previous analysis in which we had studied the dependence 
of dynamo action on the rate of SN explosions. We have relied upon 
the standard test-fields method to compute the dynamo coefficients 
in both models and simulated a simple 1-D dynamo model to examine 
the effect of various turbulent transport parameters on the inclusion 
of CR energy and growth of mean magnetic field. We have compared this with 
our previous analysis of models that do not involve the CR component, 
however for the current work we have restricted the analysis to the 
initial kinematic growth phase of the magnetic field.

One of the principle distinctions in the models that 
involved the CR was the faster growth rates of the 
mean magnetic fields, compared to models with no CR 
component. It was also found that for a model that 
involved SN explosions with purely a CR energy 
injected in the ISM, the magnetic energy had a 
slightly faster growth rate compared to the model 
that includes SN with CR and thermal energy. 
Inclusion of CR was found also to have a distinctive 
impact on the off-diagonal elements of the $ \alpha$ 
tensor. Specifically for a model which included only 
the CR, the magnitude of $\alpha_{yx}$ was found to 
be negligible compared to that of $\alpha_{xy}$. On 
the other hand the models that include the SN that 
expel both CR and thermal energy, $\alpha_{yx}$ was 
not negligible although smaller than $\alpha_{xy}$. 
This is in contrast with the purely thermal SN models 
that we had simulated earlier \citep{bendre_15}, where $\alpha_{xy}$ and $\alpha_{yx}$ were found to be statistically equal in magnitude but with opposite directions. This resulted into a systematic turbulent pumping (or $\gamma$) effect. The presence of $\gamma$ effect, is a robust result which has been seen in other previous ISM simulations as well \cite{gressel_08}.
The $\gamma$ effect arises 
as an advection in the direction of the gradient of 
turbulent intensity, which in the case of galaxies 
is towards the galactic midplane. Turbulent pumping 
effect therefore acts essentially in opposition to 
the advection governed by the outward wind. For the 
models including the CR component however this effect 
turned out to be nonuniform for $x$ and $y$ components 
of the mean magnetic field, implying the more effective 
outward transport of the $x$ component of the mean 
magnetic field compared to the $y$. Anisotropy of 
turbulent pumping is more pronounced in the 
purely CR model \verb|CR| than in model \verb|CR_TH|, 
where the isotropically propagating thermal energy is 
possibly introducing a non vanishing $\alpha_{yx}$ 
(as opposed to CR energy which diffuses preferentially 
along the mean magnetic field lines). As a consequence 
the pitch angles seen in model \verb|CR| are negligible 
compared to the ones seen in model \verb|CR_TH|.

Another important distinction in both models involving 
the CR component is the non-negligible contribution of 
the off-diagonal component of $\eta$ tensor, $\eta_{xy} 
$ and $\eta_{yx}$, compared to that in models which do 
not involve the CR. Resulting into a transverse 
diffusive transport of the mean magnetic flux. 
Furthermore it also turned out in both of these models 
that the components $\eta_{yx}$ and $\eta_{xy}$ were 
manifestly negative and positive respectively. This is 
striking, specifically because the component $\eta_{yx}
$ in combination with rotation and shear term allows a growing 
solution to the dynamo equations, depending on the sign 
of shearing term, via the R\"adler effect. This can be 
elaborated from \eref{dynamo_eq2}, 
where negative sign of $\eta_{yx}$ in the first equation 
along with $q\,\Omega\mean{B}_x$ term in second equation 
allows the growth of $\mean{\mathbf{B}}$ components, even 
in the absence of other source terms arising from 
turbulent helicity i.e. $\alpha_{i j}$. This effect was 
consistently absent in the models without the CR component, 
where both $\eta_{xy}$ and $\eta_{yx}$ were approximately 
zero within $1-\sigma$ confidence interval, even 
irrespective of the SN explosion rates. 
However, it should be noted here that the contribution 
from shear related higher order effects in the 
expansion of turbulent diffusivity tensor, such as the 
shear-current effect $\kappa$ \citep[eg.][]{rogachevskii,axel_radler} 
also gets added to the off-diagonal terms of $\eta$ tensor 
mixing up with the contributions arising from $\delta
\times\overline{\mathbf{J}}$ effect described by \cite{radler69}, 
where the $\delta$ is an expansion coefficient and $\overline{
\mathbf{J}}$ is the mean current \citep[see also the Section 4.4.3 of] []{rincon_review}. It appears impossible to
disentangle the individual contributions from the available 
data of the simulations. The presence of R\"adler effect may 
be relevant in explaining the enhanced growth rate of mean 
field expected for the dynamo operating in turbulence driven 
by the CR (see eg. \fref{fig:shear_current}), especially 
given the fact that this effect does not vanish even within 
the model that involves both CR and thermal energy inputs 
from SN explosions. 

It 
however still remains to be seen, how these results scale 
with the magnetic field aligned diffusion coefficients of 
CR energy, which we plan to address in the future. 
Nevertheless the computed sets of dynamo tensors reproduce 
the entire evolution of mean magnetic field seen in the 
DNS, using 1-D dynamo simulations as can be seen from 
\fref{fig:by_ene} and \fref{fig:b_contour_cr}. It 
would also be of our interest to analyze the saturation 
mechanism of such a dynamo where CRs are included, and to 
see how the mean field quenches the dynamo coefficients, similar to \cite{gressel2013} for the simulations without the CR. 
Another open question which we have not analyzed 
here is the one of how the CR component affects 
the multiphase structure of ISM, and how and whether
they differ characteristically from that in the simulations of ISM without the CRs \citep[eg.][]{nonlocality}.
It should however be noted here that this prescription may 
not completely describe the CR propagation in the ISM when 
dynamical strength magnetic fields exist, and gives rise to 
a CR streaming instability. Wherein the CR component 
interacts with Alfv\'en waves that are generated by itself, 
this further limits its propagation speed. This effect will 
probably be important in the dynamical phase. 
\section{Data Availability}
The data underlying this article are available in the repository "On the Combined Role of Cosmic Rays and Supernova-Driven Turbulence for Galactic Dynamos", at 
\cite{bendre_abhijit_b_2020_3992802}
%Furthermore the inclusion of CR also led to the nonlinear vertical 
%profiles of outward wind, flared in the outer halo part of the box, 
%in contrast to the purely thermal SN explosion models presented in 
%our previous analysis. A careful parameter space analysis using a 
%1-D dynamo model reveled that the combination of these two effects 
%leads the distinctive double peaked geometry of mean-field seen in 
%both models involving CR component. 
\section*{Acknowledgements}
We used the NIRVANA code version 3.3, developed 
by Udo Ziegler at the Leibniz-Institut f\"ur 
Astrophysik Potsdam (AIP). For computations we 
used Leibniz Computer Cluster, also at AIP. We 
thank K. Subramanian and Nishant Singh for very 
insightful discussions and valuable inputs.

%%%%%%%%%%%%%%%%%%%%%%%%%%%%%%%%%%%%%%%%%%%%%%%%%%



\bibliographystyle{mnras}
\bibliography{main} % if your bibtex file is called example.bib


%%%%%%%%%%%%%%%%%%%%%%%%%%%%%%%%%%%%%%%%%%%%%%%%%%

%%%%%%%%%%%%%%%%% APPENDICES %%%%%%%%%%%%%%%%%%%%%

\appendix

%%%%%%%%%%%%%%%%%%%%%%%%%%%%%%%%%%%%%%%%%%%%%%%%%%

% Don't change these lines
\bsp	% typesetting comment
\label{lastpage}
\end{document}